\documentclass[prb,nopacs,twocolumn,preprintnumbers,amsmath,amssymb]{revtex4}

\usepackage{epsfig}
\usepackage{graphicx}
\usepackage{dcolumn}
\usepackage{color}
\usepackage{bm}

\begin{document}

\title{Band structure and gaps of triangular graphene superlattices}
\author{F. Guinea}
{\affiliation{Instituto de Ciencia de Materiales de Madrid. CSIC.
Sor Juana In\'es de la Cruz 3. 28049 Madrid. Spain}
\author{Tony Low}
\affiliation{Hall for Discovery Learning Research, Purdue
University, West Lafayette, IN47907-1791 Indiana, US}

\begin{abstract}
General properties of long wavelength triangular graphene
superlattice are studied. It is shown that Dirac points with and
without gaps can arise at a number of high symmetry points of the
Brillouin Zone. The existence of gaps can lead to insulating
behavior at commensurate fillings. Strain and magnetic superlattices
are also discussed.
\end{abstract}

\maketitle \section{Introduction} Graphene is a two dimensional
metal when carriers are induced by an electric
field\cite{Netal04,Netal05,GN07,NGPNG09}. A gap at the Fermi level
has been observed by STM measurements\cite{LLA09,Zetal08} (see
also\cite{WGLB08}). We analyze the gaps induced by a periodic
structure, and the possibility that these gaps are generated
spontaneously.

Graphene superlattices have been observed in graphene layers grown
on transition metals\cite{Vetal08,Betal10} (see
also\cite{ON07,NBFM06,MGW07,Metal08,Petal08,JDD08,Uetal08}).
Superlattices are also found in graphene grown by the decomposition
of SiC\cite{Zetal07}. In general, graphene superlattices can have
interesting properties, such as highly anisotropic transport
properties\cite{PYSCL08}, or Dirac points at finite
energies\cite{PYSCL08b,TS09,BF09,BVP10,ABFKZ10}. In general, the
study of the properties of graphene superlattices has attracted
great interest, due to the many novel features their electronic
spectra can
show\cite{S09,BFSN09,AEAT09,SAL09,Getal09,Retal09,GKG10,PSYCL10,SCN10}.
In the following, we analyze general properties of the spectra of
graphene superlattices with a two dimensional triangular
periodicity. These supperlattices share the symmetries of the
graphene lattice, and are commonly found in graphene layers grown on
metallic surfaces. As discussed below, these superlattices can show
a gap at the Fermi energy for a number of commensurate fillings. It
seems likely that they can be formed spontaneously on very uniform
substrates, such as BNi, or due to intrinsic instabilities of
graphene.

In this paper, we study the general properties of triangular
graphene superlattices created by a scalar potential, followed by a
discussion of strain and magnetic superlattices.

\section{Brillouin Zone of triangular graphene superlattices} We define
the lattice vectors of the graphene lattice as:
\begin{align}
\vec{a}_1 &\equiv {\bf n}_x \nonumber \\
\vec{a}_2 &\equiv \frac{1}{2} {\bf n}_x + \frac{\sqrt{3}}{2} {\bf
n}_y
\end{align}
A triangular superlattice is described by the unit vectors:
\begin{align}
\vec{b}_1 &\equiv n_1 \vec{a}_1 + n_2 \vec{a}_2 \nonumber \\
\vec{b}_2 &\equiv - n_2 \vec{a}_1 + ( n_1 + n_2 ) \vec{a}_2
\end{align}
where $n_1$ and $n_2$ are arbitrary integers different from zero.

There are three types of high symmetry points in the Brillouin Zone
of a triangular lattice, $\Gamma , M$ and $K$. There are two
inequivalent $K$ points,  $K$ and $K'$, at the corners of the
hexagonal Brillouin Zone, and three inequivalent $M$ points, at the
centers of the edges. Time reversal exchanges $K$ and $K'$, while
leaving the $\Gamma$ point and the three $M$ points
unchanged\cite{MGV07}. The vectors which define these points are
such that:
\begin{align}
\vec{\Gamma} \vec{a}_1 &= \vec{\Gamma} \vec{a}_2 = 0 \nonumber \\
\vec{K} \vec{a}_1 &= \frac{4 \pi}{3} \, \, \, \, \, \vec{K}
\vec{a}_2 = \frac{2 \pi}{3} \nonumber \\
\vec{K}' \vec{a}_1 &= \frac{2 \pi}{3} \, \, \, \, \, \vec{K}'
\vec{a}_2 = \frac{4 \pi}{3} \nonumber \\
\vec{M}_1 \vec{a}_1 &= \pi \, \, \, \, \, \vec{M}_1 \vec{a}_2 = \pi
\nonumber \\
\vec{M}_2 \vec{a}_1 &= \pi \, \, \, \, \, \vec{M}_2 \vec{a}_2 = 0
\nonumber \\
\vec{M}_3 \vec{a}_1 &= 0 \, \, \, \, \, \vec{M}_3 \vec{a}_2 = \pi
\end{align}

\begin{figure}[!t]
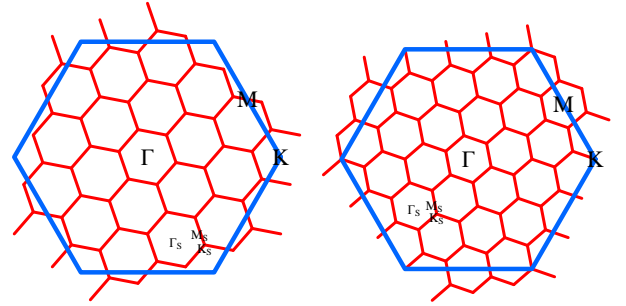

\begin{center}
\includegraphics[width=4cm,angle=0]{BZ_4_1.dat}
\includegraphics[width=4cm,angle=0]{BZ_4_2.dat}
\caption[fig]{(Color online). Examples of Brillouin Zones of
superlattices. Left: $n_1 = 1 , n_2 = 4$. Right: $n_1 = 2 , n_2 =
4$.} \label{BZ}
\end{center}
\end{figure}
The low energy states of graphene lie close to the $K$ and $K'$ of
the original Brillouin Zone. The positions of these points in the
superlattice Brillouin Zone is determined by:
\begin{align}
\vec{K}_S \vec{b}_1 &= \frac{4 \pi}{3} n_1 + \frac{2 \pi}{3} n_2 \,
\, \, \, \,  \vec{K}_S \vec{b}_2  = - \frac{4 \pi}{3} n_2 + \frac{2
\pi}{3}
( n_1 + n_2 ) \nonumber \\
\vec{K}'_S \vec{b}_1 &= \frac{2 \pi}{3} n_1 + \frac{4 \pi}{3} n_2 \,
\, \, \, \,  \vec{K}'_S \vec{b}_2  = - \frac{2 \pi}{3} n_2 + \frac{4
\pi}{3} ( n_1 + n_2 )
\end{align}
Thus, when $2 n_1 + n_2$ is a multiple of three the graphene $K$ and
$K'$ points will be mapped onto the $\Gamma_S$ point of the
superlattice Brillouin Zone. Otherwise, they will be mapped onto the
corners of the Brillouin zone, $K_S$ and $K'_S$. Examples of
superlattice Brillouin Zones are given in Fig.~\ref{BZ}.

\section{Dispersion near high symmetry points}
\subsection{The model}
We  study superlattices induced by a modulation of the on site
energy of the $\pi$ orbitals. We assume that it is a weak
perturbation of the graphene Dirac equation, except in cases where
degeneracies occur.  We consider the Fourier components of the
potential with lowest wavevector, of modulus:
\begin{equation}
G = \frac{4 \pi}{a \sqrt{3 ( n_1^2 + n_2^2 + n_1 n_2 )}}
\end{equation}
We write the potential as the sum of s symmetric part $V_G$, and an
antisymmetric part, $\Delta_G$, with respect to the interchange of
sublattices. We neglect for the moment short wavelength components
which mix the two inequivalent Dirac points of the unperturbed
graphene layer.

We analyze the changes in the Fermi velocity near the Dirac energy
induced by the superlattice potential, and the points in the lowest
bands of the superlattice where degeneracies persist when $V_G \ne
0$ and $\Delta_G = 0$. As discussed below, this situation gives rise
to a new set of Dirac equations at finite energies.

\subsection{Dirac energy at the $\Gamma_S$ point}
We consider the case when the $K$ and $K'$ points of the graphene
Brillouin Zone are mapped onto the $\Gamma_S$ point of the
superlattice Brillouin Zone. Using lowest order perturbation theory,
we find, near the $\Gamma_S$ point a renormalization of the Fermi
velocity\cite{PYSCL08,PYSCL08b}:
 \begin{equation}
 \delta v_F \approx - \frac{6 | V_G |^2}{v_F G^2} + \frac{6 | \Delta_G |^2}{v_F G^2}
 \label{v_ren}
 \end{equation}
 There is a twofold degeneracy at the three $M_S$ points, if $\Delta_G = 0$.
 The energy of these states is $v_F G / 2$. At finite distances from the $M_S$ points, we can
write an effective hamiltonian:
\begin{equation}
H \equiv \left( \begin{array}{cc} \frac{v_F G}{2} + 2 v_F k_x &
\Delta_G + \frac{4 i V_G k_y}{G} \\ \Delta_G - \frac{4 i V_G k_y}{G}
&\frac{v_F G}{2} - v_F k_x \end{array} \right) \label{dirac_m}
\end{equation}
which gives an anisotropic Dirac equation with a gap:
\begin{equation}
\epsilon_{M_S} \approx \frac{v_F G}{2} \pm \sqrt{4 v_F^2 k_x^2 +
\Delta_G^2 + \frac{16 V_G^2 k_y^2}{G^2}}
\end{equation}

At the $K_S$ and $K'_S$ points, there are three degenerate levels
for $V_G = \Delta_G = 0$, with energy $\epsilon_{K_S} = v_F G /
\sqrt{3}$. When $V_G \ne 0$ these three levels are split into a
doublet, with energy $\epsilon_{K_S}^d = v_F G / \sqrt{3} + V_G /
2$, and a singlet, at $\epsilon_{K_S}^s = v_F G / \sqrt{3} - V_G$.
Expanding around the $K_S$ point, the effective hamiltonian for the
doublet is:
\begin{equation}
H \equiv \left( \begin{array}{cc} \frac{v_F G}{\sqrt{3}} +
\frac{V_G}{2} - \frac{v_F k_x}{2} &\frac{v_F k_y}{2} - \frac{3 i
\Delta_G}{4} \\ \frac{v_F k_y}{2} + \frac{3 i \Delta_G}{4}
&\frac{v_F G}{\sqrt{3}} + \frac{V_G}{2} + \frac{v_F k_x}{2}
\end{array} \right)
\label{dirac_k}
\end{equation}
This is the two dimensional Dirac equation with a mass term. The
dispersion relation is:
\begin{align}
\epsilon_{K_S}^s &\approx \frac{v_F G}{\sqrt{3}} - V_G + O \left(
\frac{\Delta_G^2}{V_G} , \frac{v_F^2 ( k_x^2 + k_y^2 )}{V_G} \right)
\nonumber
\\
\epsilon_{K_S}^d &\approx \frac{v_F G}{\sqrt{3}} + \frac{V_G}{2} \pm
\sqrt{\frac{v_F^2 ( k_x^2 + k_y^2 )}{4} + \frac{9 \Delta_G^2}{16}}
\end{align}
There are two sets of degenerate bands, derived from the $K$ and
$K'$ points of the Brillouin Zone of graphene. This degeneracy will
be broken by short wavelength terms in the superlattice potential.

\subsection{Dirac energy at the $K_S$ and $K'_S$ points}
The renormalization of the Fermi velocity near the $K_S$ and $K'_S$
points is the same as in eq.~\ref{v_ren}.

\begin{figure}[!t]
\begin{center}
\includegraphics[width=4cm,angle=0]{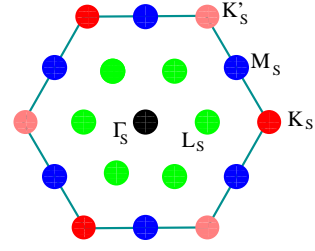}
\caption[fig]{(Color online). Points in the superlattice Brillouin
Zone where degeneracies occur when $V_G \ne 0$.} \label{BZ_K}
\end{center}
\end{figure}
There are doubly degenerate states, even when $V_G \ne 0$, at the
six inequivalent points at positions $\vec{L}_S = \vec{K}_S / 2$, as
shown in Fig.~\ref{BZ_K}. The energy of these points is
$\epsilon_{L_S} = v_F G / 2$. Expanding around these points, we find
an effective anisotropic Dirac equation, given by
eq.(\ref{dirac_m}).

There are another set of doubly degenerate states at the $M_S$
points. The two states arise from the $K$ and $K'$ points of the
original graphene Brillouin Zone. The degeneracy persists when $V_G
\ne 0$ and $\Delta_G \ne 0$, and is only broken by short wavelength
components of the superlattice potential. When these components are
finite, an effective anisotropic Dirac equation will arise similar
to that in eq.(\ref{dirac_m}).

For $V_G = \Delta_G = 0$ there are six degenerate states at the
$\Gamma_S$ point. The long range part of the superlattice potential
will hybridize states which are derived from the $K$ and $K'$ points
of the original graphene Brillouin Zone. We obtain two sets of
isotropic Dirac equations, described by eq.(\ref{dirac_k}), and two
degenerate states. The short range part of the Dirac equation will
break these degeneracies.
\begin{figure}[!t]
\begin{center}
\includegraphics[width=8cm,angle=0]{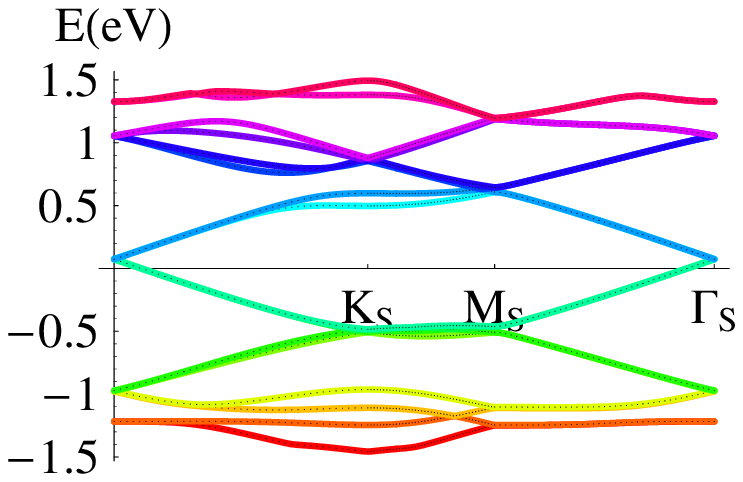}
\includegraphics[width=4cm,angle=0]{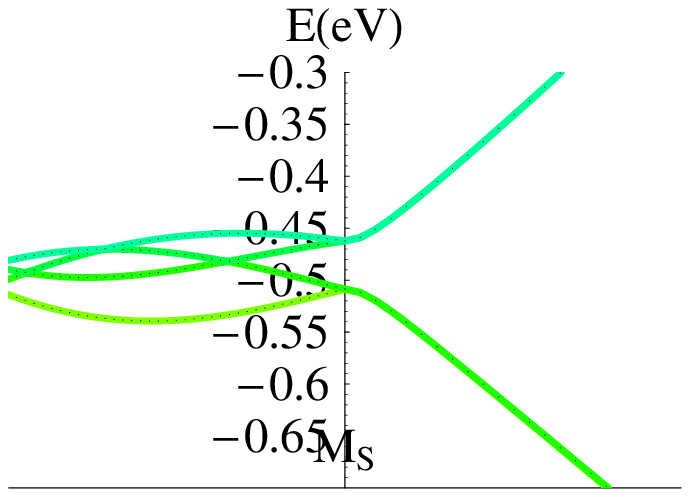}
\includegraphics[width=4cm,angle=0]{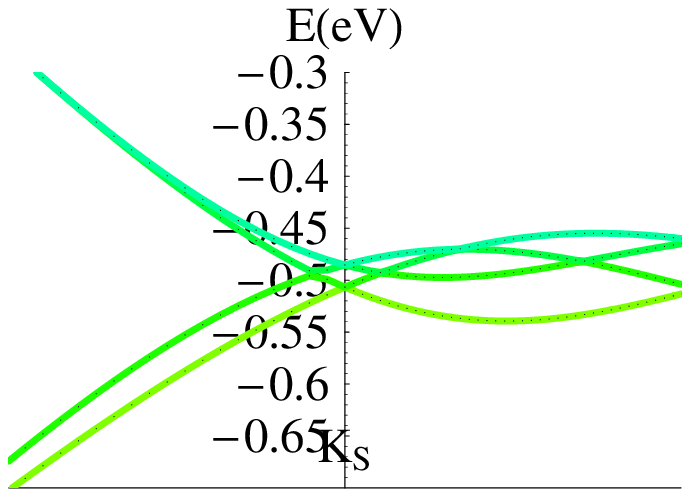}
\caption[fig]{(Color online). Top: Low energy bands for a $12 \times
12$ superlattice, with $V_G = 0.3$eV and $\Delta_G = 0$. Bottom:
Left: Details near the $M_S$ point. Right: Detail near the $K_S$
point.} \label{bands_12_0.3_0}
\end{center}
\end{figure}

\begin{figure}[!t]
\begin{center}
\includegraphics[width=8cm,angle=0]{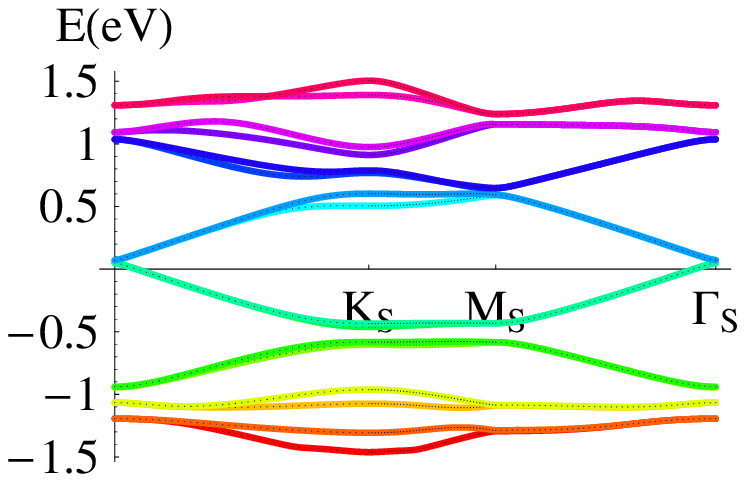}
\includegraphics[width=4cm,angle=0]{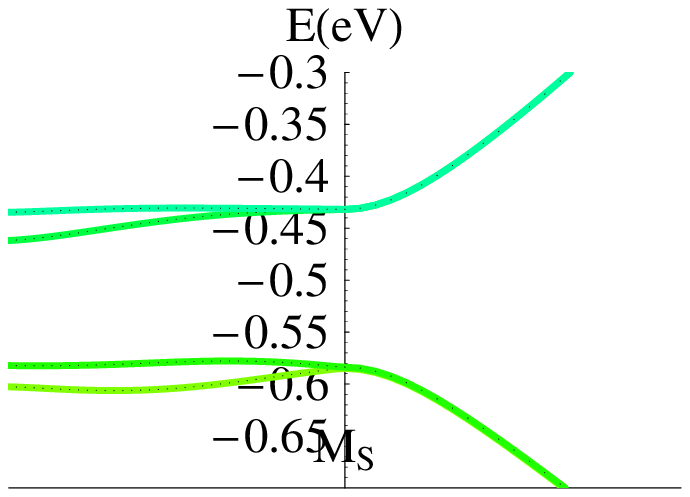}
\includegraphics[width=4cm,angle=0]{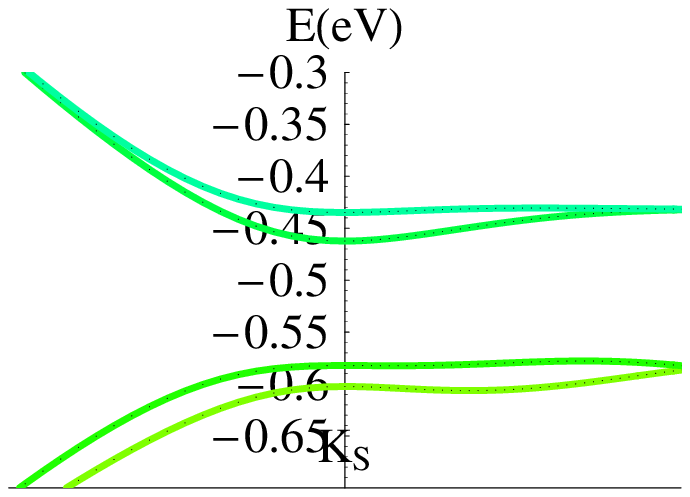}
\caption[fig]{(Color online). Top: Low energy bands for a $12 \times
12$ superlattice, with $V_G = 0.3$eV and $\Delta_G = 0.1$eV. Bottom:
Left: Details near the $M_S$ point. Right: Detail near the $K_S$
point.} \label{bands_12_0.3_0.1}
\end{center}
\end{figure}

\begin{figure}[!t]
\begin{center}
\includegraphics[width=8cm,angle=0]{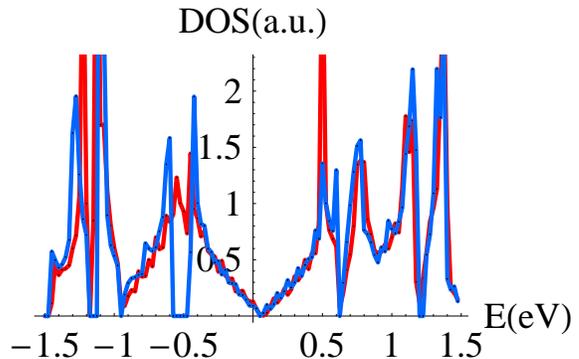}
\caption[fig]{(Color online). Density of states for a $12 \times 12$
superlattice. Red: $V_G = 0.3$eV and $\Delta_G = 0$. Blue: $V_G =
0.3$eV and $\Delta_G = 0.1$eV.} \label{dos}
\end{center}
\end{figure}
\subsection{Results} We analyze the bands induced by a  $N \times N$
superlattice. The hopping matrix between $\pi$ orbitals in
neighboring carbon atoms is $t = 3$eV. The bands for $V_G = 0.3$eV
and $\Delta_G = 0$ are shown in Fig.~\ref{bands_12_0.3_0}. The bands
show Dirac points at the $M_S$ and $K_S$ points. When $\Delta_G$ is
increased to $\Delta_G = 0.1$eV a gap appears between successive
bands, as shown in Fig.~\ref{bands_12_0.3_0.1}. The density of
states for those two cases is shown in Fig.~\ref{dos}. Note that the
potential breaks the electron-hole symmetry of clean graphene, and
the gaps are not of the same magnitude for positive and negative
energies.

The results are in reasonable agreement with the analytical
description in the previous section. A gap of order $2 \Delta_G$ is
induced at the $M_S$ point. In order for this gap to be possible,
the following inequalities must be satisfied: \begin{align}
 \epsilon_{K_S}^s \approx \frac{v_F G}{\sqrt{3}} - V_G &\le \epsilon_{M_S}^-
 \approx \frac{v_F G}{2} - \Delta_G \nonumber \\
 \epsilon_{M_S}^+ \approx \frac{v_F G}{2} + \Delta_G &\le
 \epsilon_{K_S}^d \approx \frac{v_F G}{\sqrt{3}} + \frac{V_G}{2} -
 \frac{3 \Delta_G}{4}
 \label{inequalities}
 \end{align}
 The scaling properties of the Dirac equation imply that, if the
 dimension of the superlattice is increased, $G \rightarrow \lambda
 G$, with $\lambda < 1$, a rescaling of the superlattice potential,
 $V_G \rightarrow \lambda V_G , \Delta_G \rightarrow \lambda
 \Delta_G$, will lead to the same band structure, with energies
 scaled as $E \rightarrow \lambda E$.

\section{Strain superlattices}
A superlattice can also be produced by inducing strains, which
modulate the interatomic hoppings. The corresponding perturbation
can be seen as a gauge field, $\vec{\bf A}$, which shifts locally
the momentum\cite{VKG10}. A simple case is when the strains are due
to height modulations, $h ( \vec{\bf r} )$, which can be induced by
a substrate. In terms of the Fourier components of the modulation,
$h_{\vec{\bf G}}$, and allowing for the relaxation of the in plane
displacements, the effective gauge field can be written
as\cite{GHL08}:
\begin{widetext}
\begin{align}
A_x ( \vec{\bf G} ) &= \frac{( \lambda + \mu ) ( G_x^2 - G_y^2 )
\left[ h^{xx}_{\vec{\bf G}} G_y^2 -( h^{xy}_{\vec{\bf G}} +
h^{yx}_{\vec{\bf G}} ) G_x G_y + h^{yy}_{\vec{\bf G}} G_x^2
\right]}{| \vec{\bf G} |^4 ( \lambda + 2 \mu )} \nonumber \\
A_y ( \vec{\bf G} ) &= \frac{( \lambda + \mu ) 2 G_x G_y \left[
h^{xx}_{\vec{\bf G}} G_y^2 -( h^{xy}_{\vec{\bf G}} +
h^{yx}_{\vec{\bf G}} ) G_x G_y + h^{yy}_{\vec{\bf G}} G_x^2
\right]}{| \vec{\bf G} |^4 ( \lambda + 2 \mu )} \label{strain}
\end{align}
\end{widetext}
where the tensor $h^{ij}_{\vec{\bf G}}$ is the Fourier transform of
the functions:
\begin{align}
h^{ij} ( \vec{\bf r} ) &= \frac{\partial h}{\partial x_i}
\frac{\partial h}{\partial x_j} \label{height}
\end{align}
and $\lambda$ and $\mu$ are the elastic Lam\'e coefficients of
graphene. The field in eq.~\ref{strain} has opposite signs in the
two valleys in the Brillouin Zone.

The calculation of the effective magnetic field induced by the gauge
field in eq.~\ref{strain} is simplified when, as in the previous
sections, only one component, $h ( \vec{\bf r} ) = h_G \sum_{l = 1 ,
\cdots , 6}  e^{i \vec{\bf G}_l \vec{\bf r}}$,  in a superlattice is
considered. The tensor in eq.~\ref{height} has non zero components
for all combinations of the type $\vec{\bf G}_k + \vec{\bf G}_l$.
When the gauge field, $\vec{\bf A} ( \vec{\bf G} )$ is parallel to
$\vec{\bf G} )$, the vector potential can be gauged away, and does
not induce an effective magnetic field. This implies that out of the
18 possible values of the vector $\vec{\bf G}_k + \vec{\bf G}_l$,
only six contribute to the effective magnetic field. These vectors
are given by $\vec{\bf G} \equiv G ( 3/2 , \sqrt{3}/2 )$ and the
vectors equivalent to it by a symmetry transformation. After some
algebra, we obtain for the effective magnetic field:
\begin{widetext}
\begin{align}
B_{strain} ( \vec{\bf r} ) &= \frac{\beta}{a} \frac{\lambda +
\mu}{\lambda + 2 \mu} \frac{27 \sqrt{3}}{8} G^3 h_G^2 \left[ 2 \cos
( \sqrt{3} G y ) + 4 \cos \left( \frac{\sqrt{3} G y}{2} \right)
\left( \frac{3 G x}{2} \right) \right] \label{field}
\end{align}
\end{widetext}
where $\beta = \partial \log ( t ) / \partial \log ( a ) \approx 2 -
3$, $t \approx 3$eV is the nearest neighbor hopping, and $a \approx
1.4$\AA \, is the distance between nearest neighbor carbon atoms.
The superlattice defined by the effective magnetic field has a unit
vector of length $\sqrt{3} G$, so that the area of its unit cell is
a smaller than the area of the unit cell of the original
superlattice by a factor $1/3$.

Using eq.~\ref{field}, and $G = 2 \pi / L$, where $L$ is the length
of the unit vector of the superlattice, we can write the magnetic
length associated to the maximum effective field in the system as:
\begin{align}
\frac{1}{\ell_B^2} &= \frac{\lambda+\mu}{\lambda+2\mu}
\frac{\beta}{a} \frac{9 \sqrt{3} \pi^3}{2} \frac{h_{max}^2}{L^3}
\end{align}
where $h_{max} = 6 h_G$ is the maximum value of $h ( \vec{\bf G} )$,
assuming $h_G > 0$. For values $h_{max} \approx 1$nm, and $L \approx
40$nm, we find $l_B \approx 14$nm, so that the effective field is
such that $B_{strain}^{max} \approx 1 - 2$T.

\section{Magnetic superlattices}
A superlattice can also be induced by a spatially modulated magnetic
field. A combination of a modulated magnetic field and a scalar
potential opens a gap at the Dirac energy, and the resulting
insulator is a Quantum Hall system\cite{S09}, with chiral currents
at the boundaries\cite{H88}. Here we obtain this effect using second
order perturbation theory, instead of the arguments used
in~\cite{S09}. As in the previous sections, we assume the simplest
periodicity compatible with the superlattice hexagonal symmetry
\begin{align}
V ( \vec{\bf r} ) &=   V_G \sum_{l=1,\cdots,6} e^{i \vec{\bf G}_l
\vec{\bf r}} \nonumber \\
B ( \vec{\bf r} ) &= B_G \sum_{l=1,\cdots,6} e^{i \vec{\bf G}_l
\vec{\bf r}} \label{magnetic}
\end{align}
The eigenstates of the unperturbed hamiltonian at the $K$ and $K'$
points of the Brillouin Zone can be written as $| 0 \rangle_A = ( 1
, 0 )$ and $| 0 \rangle_B = ( 0 , 1 )$, which each component of the
spinor corresponds to one sublattice. These states are hybridized
with states $| \vec{\bf G}_\pm \rangle_K = ( 1 , \pm e^{i
\phi_{\vec{\bf G}}} )$  and $| \vec{\bf G}_\pm \rangle_{K'} = ( 1 ,
\mp e^{- i \phi_{\vec{\bf G}}} )$, with energies $\epsilon_{\vec{\bf
G}} = \pm v_F | \vec{\bf G} |$, and $e^{ i \phi_{\vec{\bf G}}} = (
G_x + i G_y ) / | \vec{\bf G} |$.

The energy of states $| 0 \rangle_A$ and $| 0 \rangle_B$ are
modified in different ways by virtual hoppings into states $|
\vec{\bf G}_\pm \rangle_K$ and $| \vec{\bf G}_\pm \rangle_{K'}$,
leading to gaps in both valleys. Moreover, the gaps have different
signs, showing that time reversal symmetry in the system is broken,
and that a Quantum Hall phase has been induced. The gap can be
written as
\begin{align}
\Delta &= \pm \sum_{l=1,\cdots,6} \frac{2 v_F {\rm Re} \left\{
V_{\vec{\bf G}_l}^* \left[ A_x ( \vec{\bf G}_l ) + i A_y ( \vec{\bf
G}_l ) \right] e^{- i \phi_{\vec{\bf G}_l}} \right\}}{\left|
\epsilon_{\vec{\bf G}_l} \right|}
\end{align}
where $\vec{\bf A}_{\vec{\bf G}}$ is the vector potential, which we
define as:
\begin{align}
A_x ( \vec{\bf G} ) &= \frac{i G_y}{| \vec{\bf G} |^2}
\frac{B_{\vec{\bf G}}}{\Phi_0}
\nonumber \\
A_y ( \vec{\bf G} ) &=  \frac{-i G_x}{| \vec{\bf G} |^2}
\frac{B_{\vec{\bf G}}}{\Phi_0}
\end{align}
where $\Phi_0 = e h / c$ is the quantum unit of flux. Using this
expression, we finally obtain:
\begin{align}
\Delta &= 12 \frac{B_G V_G}{\Phi_0 G^2}
\end{align}
in agreement with\cite{S09}.
\section{Self consistent opening of a gap}
The previous analysis shows that gap can open at finite energies in
graphene in the presence of a superlattice potential with a
staggered component. When the number of carriers is such that only a
small number of subbands are completely filled and the rest are
completely empty the electronic energy will be lowered in the
presence of the gap. A lattice distortion which leads to the
appropriate potential will be energetically favorable if the gain in
electronic energy exceeds the formation energy of the distortion, as
in the Peierls instability in one dimension.

In graphene on top of a metal or other substrate with a large
dielectric constant, as in\cite{Vetal08}, out of plane displacements
lead to changes in the on site energies of the $\pi$ orbitals. An
electron in a given carbon atom experiences the image potential due
to the screening. A change of position of $\Delta z$ leads to a
change of the image potential of order $e e^* \Delta z / ( 4 d^2 )$,
where $e^* = e ( \epsilon_0 - 1 ) / ( \epsilon_0 + 1)$ is the image
charge, $\epsilon_0$ is the dielectric constant of the substrate,
and $d$ is the distance to the substrate. A vertical displacement of
$\Delta z \sim 1$\AA \, when the graphene layer is at a distance $d
\approx 3$\AA \, of the substrate can lead to shifts of the onsite
energies of order 0.1eV. The electronic gain of energy due to the
existence of a gap, per unit cell, is then $E_{elec} \approx
\Delta_G \approx e e^* \Delta z d^{-2}$.

The elastic energy per unit cell required to create a staggered
distortion of amplitude $\Delta z$ is of order $E_{elas} \approx
\kappa \Delta z^2 a^{-2}$, where $\kappa \approx 1$eV is the bending
rigidity of graphene.

A gap will exist above a threshold for the superlattice potential,
$V_G \sim \Delta_G \gtrsim v_F G ( 1 / \sqrt{3} - 1 / 2 )$. The area
of the Brillouin Zone of the supercell, $\sqrt{3} G^2 / 2$ should be
close to the area within the Fermi surface of the unperturbed
graphene, $\pi k_F^2$. Hence, a spontaneous staggered distortion is
favored if:
\begin{align}
E_{elec} + E_{elas} &\approx - \frac{e e^* \Delta z}{d^2} + \kappa
\left( \frac{\Delta z}{a} \right)^2 < 0 \nonumber \\
V_G &\approx \frac{e e^* \Delta z}{d^2} > v_F G \left(
\frac{1}{\sqrt{3}} - \frac{1}{2} \right) \nonumber \\
\frac{\sqrt{3} G^2}{2} &\approx \pi k_F^2 \label{inequalities_2}
\end{align}
The last equation in (\ref{inequalities_2}) implies that $G \propto
k_F$, and from the first two we obtain that $v_F G \lesssim ( e e^*
)^2 a^2 d^{-4} \kappa^{-1}$. Hence, a spontaneous distortion is
energetically favored for carrier densities such that:
\begin{equation}
\rho = \pi k_F^2 \lesssim \frac{ ( e e^* )^4 a^4}{v_F^2 \kappa^2
d^8}
\end{equation}
and the staggered distortion, with inverse wavelength $G \sim k_F$
of the order:
\begin{equation}
\Delta z \sim \frac{v_F k_F d^2}{e e^*}
\end{equation}
For a metal, we have $e_0 = \infty$ and $e^* = e$. For $a \approx
2$\AA, $d \approx 3$\AA \, and $\kappa \approx 1$eV, we find that a
staggered corrugation is energetically favored for carrier densities
$\rho \lesssim 10^{13}$cm$^{-2}$, leading to maximum deformations of
0.2 \AA. For SiO$_2$, where $\epsilon_0 = 3.9$ and $e^* \approx 0.6
e$, the corrugations takes place for carrier densities $\rho
\lesssim 3 \times 10^{12}$cm$^{-2}$, and maximum deformations of
0.15 \AA.

The formation of these long wavelength modulations is only possible
in systems with a high degree of order, as the superlattice features
will be reduced by disorder. For instance, weak scatterers with
concentration $n_{imp}$, which change the onsite potential by an
amount of order $\Delta_G$, lead to a mean free path $l_{el} \sim
v_F^2 / ( \Delta_G^2 k_F a^4 n_{imp} )$. A large concentration of
these defects, $n_{imp} \sim a^{-2}$ will suppress the formation of
a superlattice, while affecting little the conductivity, $\sigma
\sim ( e^2 / h ) \times ( k_F l_{el} ) \sim
 ( e^2 / h ) \times ( v_F^2 a^{-2} \Delta_G^{-2} )$.

\section{Conclusions}
We have analyzed the formation of Dirac points and gaps at high
energy points of triangular graphene superlattices. We have shown
that in some cases a gap can be formed over the entire Fermi
surface, making graphene insulating. We have discussed instabilities
which might give rise to the spontaneous formation of gaps of this
type. General properties of strain and magnetic superlattices are
also discussed.

\section{Acknowledgements}
We appreciate useful conversations with J. L. Ma\~nes, R. Miranda,
and A. V\'azquez de Praga. Funding from MICINN (Spain) through
grants FIS2008-00124 and CONSOLIDER CSD2007-00010 is gratefully
acknowledged. TL acknowledges funding from INDEX/NSF (US).
\bibliography{superlattice}

\begin{thebibliography}{36}
\expandafter\ifx\csname natexlab\endcsname\relax\def\natexlab#1{#1}\fi
\expandafter\ifx\csname bibnamefont\endcsname\relax
  \def\bibnamefont#1{#1}\fi
\expandafter\ifx\csname bibfnamefont\endcsname\relax
  \def\bibfnamefont#1{#1}\fi
\expandafter\ifx\csname citenamefont\endcsname\relax
  \def\citenamefont#1{#1}\fi
\expandafter\ifx\csname url\endcsname\relax
  \def\url#1{\texttt{#1}}\fi
\expandafter\ifx\csname urlprefix\endcsname\relax\def\urlprefix{URL }\fi
\providecommand{\bibinfo}[2]{#2}
\providecommand{\eprint}[2][]{\url{#2}}

\bibitem[{\citenamefont{Novoselov et~al.}(2004)\citenamefont{Novoselov, Geim,
  Morozov, Jiang, Zhang, Dubonos, Grigorieva, , and Firsov}}]{Netal04}
\bibinfo{author}{\bibfnamefont{K.~S.} \bibnamefont{Novoselov}},
  \bibinfo{author}{\bibfnamefont{A.~K.} \bibnamefont{Geim}},
  \bibinfo{author}{\bibfnamefont{S.~V.} \bibnamefont{Morozov}},
  \bibinfo{author}{\bibfnamefont{D.}~\bibnamefont{Jiang}},
  \bibinfo{author}{\bibfnamefont{Y.}~\bibnamefont{Zhang}},
  \bibinfo{author}{\bibfnamefont{S.~V.} \bibnamefont{Dubonos}},
  \bibinfo{author}{\bibfnamefont{I.~V.} \bibnamefont{Grigorieva}}, ,
  \bibnamefont{and} \bibinfo{author}{\bibfnamefont{A.~A.}
  \bibnamefont{Firsov}}, \bibinfo{journal}{Science}
  \textbf{\bibinfo{volume}{306}}, \bibinfo{pages}{666} (\bibinfo{year}{2004}).

\bibitem[{\citenamefont{Novoselov et~al.}(2005)\citenamefont{Novoselov, Jiang,
  Schedin, Booth, Khotkevich, Morozov, and Geim}}]{Netal05}
\bibinfo{author}{\bibfnamefont{K.~S.} \bibnamefont{Novoselov}},
  \bibinfo{author}{\bibfnamefont{D.}~\bibnamefont{Jiang}},
  \bibinfo{author}{\bibfnamefont{F.}~\bibnamefont{Schedin}},
  \bibinfo{author}{\bibfnamefont{T.~J.} \bibnamefont{Booth}},
  \bibinfo{author}{\bibfnamefont{V.~V.} \bibnamefont{Khotkevich}},
  \bibinfo{author}{\bibfnamefont{S.~V.} \bibnamefont{Morozov}},
  \bibnamefont{and} \bibinfo{author}{\bibfnamefont{A.~K.} \bibnamefont{Geim}},
  \bibinfo{journal}{Proc. Natl. Acad. Sci. U.S.A.}
  \textbf{\bibinfo{volume}{102}}, \bibinfo{pages}{10451}
  (\bibinfo{year}{2005}).

\bibitem[{\citenamefont{Geim and Novoselov}(2007)}]{GN07}
\bibinfo{author}{\bibfnamefont{A.~K.} \bibnamefont{Geim}} \bibnamefont{and}
  \bibinfo{author}{\bibfnamefont{K.~S.} \bibnamefont{Novoselov}},
  \bibinfo{journal}{Nature Materials} \textbf{\bibinfo{volume}{6}},
  \bibinfo{pages}{183} (\bibinfo{year}{2007}).

\bibitem[{\citenamefont{{Castro Neto} et~al.}(2009)\citenamefont{{Castro Neto},
  Guinea, Peres, Novoselov, and Geim}}]{NGPNG09}
\bibinfo{author}{\bibfnamefont{A.~H.} \bibnamefont{{Castro Neto}}},
  \bibinfo{author}{\bibfnamefont{F.}~\bibnamefont{Guinea}},
  \bibinfo{author}{\bibfnamefont{N.~M.~R.} \bibnamefont{Peres}},
  \bibinfo{author}{\bibfnamefont{K.~S.} \bibnamefont{Novoselov}},
  \bibnamefont{and} \bibinfo{author}{\bibfnamefont{A.~K.} \bibnamefont{Geim}},
  \bibinfo{journal}{Rev. Mod. Phys.} \textbf{\bibinfo{volume}{81}},
  \bibinfo{pages}{109} (\bibinfo{year}{2009}).

\bibitem[{\citenamefont{Li et~al.}(2009)\citenamefont{Li, Luican, and
  Andrei}}]{LLA09}
\bibinfo{author}{\bibfnamefont{G.}~\bibnamefont{Li}},
  \bibinfo{author}{\bibfnamefont{A.}~\bibnamefont{Luican}}, \bibnamefont{and}
  \bibinfo{author}{\bibfnamefont{E.~Y.} \bibnamefont{Andrei}},
  \bibinfo{journal}{Phys. Rev. Lett.} \textbf{\bibinfo{volume}{102}},
  \bibinfo{pages}{176804} (\bibinfo{year}{2009}).

\bibitem[{\citenamefont{Zhang et~al.}(2008)\citenamefont{Zhang, Brar, Wang,
  Girit, Yayon, Panlasigui, Zettl, and Crommie}}]{Zetal08}
\bibinfo{author}{\bibfnamefont{Y.}~\bibnamefont{Zhang}},
  \bibinfo{author}{\bibfnamefont{V.~W.} \bibnamefont{Brar}},
  \bibinfo{author}{\bibfnamefont{F.}~\bibnamefont{Wang}},
  \bibinfo{author}{\bibfnamefont{C.}~\bibnamefont{Girit}},
  \bibinfo{author}{\bibfnamefont{Y.}~\bibnamefont{Yayon}},
  \bibinfo{author}{\bibfnamefont{M.}~\bibnamefont{Panlasigui}},
  \bibinfo{author}{\bibfnamefont{A.}~\bibnamefont{Zettl}}, \bibnamefont{and}
  \bibinfo{author}{\bibfnamefont{M.~F.} \bibnamefont{Crommie}},
  \bibinfo{journal}{Nature Phys.} \textbf{\bibinfo{volume}{4}},
  \bibinfo{pages}{627} (\bibinfo{year}{2008}).

\bibitem[{\citenamefont{Wehling et~al.}(2008)\citenamefont{Wehling, Grigorenko,
  Lichtenstein, and Balatsky}}]{WGLB08}
\bibinfo{author}{\bibfnamefont{T.~O.} \bibnamefont{Wehling}},
  \bibinfo{author}{\bibfnamefont{I.}~\bibnamefont{Grigorenko}},
  \bibinfo{author}{\bibfnamefont{A.~I.} \bibnamefont{Lichtenstein}},
  \bibnamefont{and} \bibinfo{author}{\bibfnamefont{A.~V.}
  \bibnamefont{Balatsky}}, \bibinfo{journal}{Phys. Rev. Lett.}
  \textbf{\bibinfo{volume}{101}}, \bibinfo{pages}{216803}
  (\bibinfo{year}{2008}).

\bibitem[{\citenamefont{{V\'azquez de Parga}
  et~al.}(2008)\citenamefont{{V\'azquez de Parga}, Calleja, Borca, Passeggi,
  Hinarejos, Guinea, and Miranda}}]{Vetal08}
\bibinfo{author}{\bibfnamefont{A.~L.} \bibnamefont{{V\'azquez de Parga}}},
  \bibinfo{author}{\bibfnamefont{F.}~\bibnamefont{Calleja}},
  \bibinfo{author}{\bibfnamefont{B.}~\bibnamefont{Borca}},
  \bibinfo{author}{\bibfnamefont{M.~C.} \bibnamefont{Passeggi}},
  \bibinfo{author}{\bibfnamefont{J.~J.} \bibnamefont{Hinarejos}},
  \bibinfo{author}{\bibfnamefont{F.}~\bibnamefont{Guinea}}, \bibnamefont{and}
  \bibinfo{author}{\bibfnamefont{R.}~\bibnamefont{Miranda}},
  \bibinfo{journal}{Phys. Rev. Lett.} \textbf{\bibinfo{volume}{100}},
  \bibinfo{pages}{056807} (\bibinfo{year}{2008}).

\bibitem[{\citenamefont{Borca et~al.}(2010)\citenamefont{Borca, Barja, Garnica,
  Minniti, Politano, {Rodriguez-Garc{\'\i}a}, Hinarejos, {Daniel Far{\'\i}as},
  {V\'azquez de Parga}, and Miranda}}]{Betal10}
\bibinfo{author}{\bibfnamefont{B.}~\bibnamefont{Borca}},
  \bibinfo{author}{\bibfnamefont{S.}~\bibnamefont{Barja}},
  \bibinfo{author}{\bibfnamefont{M.}~\bibnamefont{Garnica}},
  \bibinfo{author}{\bibfnamefont{M.}~\bibnamefont{Minniti}},
  \bibinfo{author}{\bibfnamefont{A.}~\bibnamefont{Politano}},
  \bibinfo{author}{\bibfnamefont{J.~M.} \bibnamefont{{Rodriguez-Garc{\'\i}a}}},
  \bibinfo{author}{\bibfnamefont{J.~J.} \bibnamefont{Hinarejos}},
  \bibinfo{author}{\bibnamefont{{Daniel Far{\'\i}as}}},
  \bibinfo{author}{\bibfnamefont{A.~L.} \bibnamefont{{V\'azquez de Parga}}},
  \bibnamefont{and} \bibinfo{author}{\bibfnamefont{R.}~\bibnamefont{Miranda}}
  (\bibinfo{year}{2010}), \eprint{arXiv:1005.1764}.

\bibitem[{\citenamefont{Oshima and Nagashima}(1997)}]{ON07}
\bibinfo{author}{\bibfnamefont{C.}~\bibnamefont{Oshima}} \bibnamefont{and}
  \bibinfo{author}{\bibfnamefont{A.}~\bibnamefont{Nagashima}},
  \bibinfo{journal}{J. Phys. Condens. Matter} \textbf{\bibinfo{volume}{9}},
  \bibinfo{pages}{1} (\bibinfo{year}{1997}).

\bibitem[{\citenamefont{N'Diaye et~al.}(2006)\citenamefont{N'Diaye, Bleikamp,
  Feibelman, and Michely}}]{NBFM06}
\bibinfo{author}{\bibfnamefont{A.~T.} \bibnamefont{N'Diaye}},
  \bibinfo{author}{\bibfnamefont{S.}~\bibnamefont{Bleikamp}},
  \bibinfo{author}{\bibfnamefont{P.~J.} \bibnamefont{Feibelman}},
  \bibnamefont{and} \bibinfo{author}{\bibfnamefont{T.}~\bibnamefont{Michely}},
  \bibinfo{journal}{Phys. Rev. Lett.} \textbf{\bibinfo{volume}{97}},
  \bibinfo{pages}{215501} (\bibinfo{year}{2006}).

\bibitem[{\citenamefont{Marchini et~al.}(2007)\citenamefont{Marchini,
  G´{\"u}nther, and Wintterlin}}]{MGW07}
\bibinfo{author}{\bibfnamefont{S.}~\bibnamefont{Marchini}},
  \bibinfo{author}{\bibfnamefont{S.}~\bibnamefont{G´{\"u}nther}},
  \bibnamefont{and}
  \bibinfo{author}{\bibfnamefont{J.}~\bibnamefont{Wintterlin}},
  \bibinfo{journal}{Phys. Rev. B} \textbf{\bibinfo{volume}{76}},
  \bibinfo{pages}{075429} (\bibinfo{year}{2007}).

\bibitem[{\citenamefont{Martoccia et~al.}(2008)\citenamefont{Martoccia,
  Willmott, Brugger, {Bj\"orck}, {G\"unther}, Schlep{\"u}tz, Crevellino, Pauli,
  Patterson, Marchini et~al.}}]{Metal08}
\bibinfo{author}{\bibfnamefont{D.}~\bibnamefont{Martoccia}},
  \bibinfo{author}{\bibfnamefont{P.~R.} \bibnamefont{Willmott}},
  \bibinfo{author}{\bibfnamefont{T.}~\bibnamefont{Brugger}},
  \bibinfo{author}{\bibfnamefont{M.}~\bibnamefont{{Bj\"orck}}},
  \bibinfo{author}{\bibfnamefont{S.}~\bibnamefont{{G\"unther}}},
  \bibinfo{author}{\bibfnamefont{C.~M.} \bibnamefont{Schlep{\"u}tz}},
  \bibinfo{author}{\bibfnamefont{A.}~\bibnamefont{Crevellino}},
  \bibinfo{author}{\bibfnamefont{S.~A.} \bibnamefont{Pauli}},
  \bibinfo{author}{\bibfnamefont{B.~D.} \bibnamefont{Patterson}},
  \bibinfo{author}{\bibfnamefont{S.}~\bibnamefont{Marchini}},
  \bibnamefont{et~al.}, \bibinfo{journal}{Phys. Rev. Lett.}
  \textbf{\bibinfo{volume}{101}}, \bibinfo{pages}{126102}
  (\bibinfo{year}{2008}).

\bibitem[{\citenamefont{Pan et~al.}(2008)\citenamefont{Pan, Jiang, Sun, Shi,
  Du, Liu, and Gao}}]{Petal08}
\bibinfo{author}{\bibfnamefont{Y.}~\bibnamefont{Pan}},
  \bibinfo{author}{\bibfnamefont{N.}~\bibnamefont{Jiang}},
  \bibinfo{author}{\bibfnamefont{J.~T.} \bibnamefont{Sun}},
  \bibinfo{author}{\bibfnamefont{D.~X.} \bibnamefont{Shi}},
  \bibinfo{author}{\bibfnamefont{S.~X.} \bibnamefont{Du}},
  \bibinfo{author}{\bibfnamefont{F.}~\bibnamefont{Liu}}, \bibnamefont{and}
  \bibinfo{author}{\bibfnamefont{H.-J.} \bibnamefont{Gao}},
  \bibinfo{journal}{Adv. Mat.} \textbf{\bibinfo{volume}{20}},
  \bibinfo{pages}{1} (\bibinfo{year}{2008}).

\bibitem[{\citenamefont{Jiang et~al.}(2008)\citenamefont{Jiang, Du, and
  Dai}}]{JDD08}
\bibinfo{author}{\bibfnamefont{D.-E.} \bibnamefont{Jiang}},
  \bibinfo{author}{\bibfnamefont{M.-H.} \bibnamefont{Du}}, \bibnamefont{and}
  \bibinfo{author}{\bibfnamefont{S.}~\bibnamefont{Dai}} (\bibinfo{year}{2008}),
  \eprint{arXiv:0901.1101}.

\bibitem[{\citenamefont{Usachov et~al.}(2008)\citenamefont{Usachov,
  Dobrotvorskii, Varykhalov, Rader, Gudat, Shikin, and Adamchuk}}]{Uetal08}
\bibinfo{author}{\bibfnamefont{D.}~\bibnamefont{Usachov}},
  \bibinfo{author}{\bibfnamefont{A.~M.} \bibnamefont{Dobrotvorskii}},
  \bibinfo{author}{\bibfnamefont{A.}~\bibnamefont{Varykhalov}},
  \bibinfo{author}{\bibfnamefont{O.}~\bibnamefont{Rader}},
  \bibinfo{author}{\bibfnamefont{W.}~\bibnamefont{Gudat}},
  \bibinfo{author}{\bibfnamefont{A.~M.} \bibnamefont{Shikin}},
  \bibnamefont{and} \bibinfo{author}{\bibfnamefont{V.~K.}
  \bibnamefont{Adamchuk}}, \bibinfo{journal}{Phys. Rev. B}
  \textbf{\bibinfo{volume}{78}}, \bibinfo{pages}{085403}
  (\bibinfo{year}{2008}).

\bibitem[{\citenamefont{Zhou et~al.}(2007)\citenamefont{Zhou, Gweon, Fedorov,
  First, {de Heer}, Lee, Guinea, Neto, and Lanzara}}]{Zetal07}
\bibinfo{author}{\bibfnamefont{S.~Y.} \bibnamefont{Zhou}},
  \bibinfo{author}{\bibfnamefont{G.-H.} \bibnamefont{Gweon}},
  \bibinfo{author}{\bibfnamefont{A.~V.} \bibnamefont{Fedorov}},
  \bibinfo{author}{\bibfnamefont{P.~N.} \bibnamefont{First}},
  \bibinfo{author}{\bibfnamefont{W.~A.} \bibnamefont{{de Heer}}},
  \bibinfo{author}{\bibfnamefont{D.-H.} \bibnamefont{Lee}},
  \bibinfo{author}{\bibfnamefont{F.}~\bibnamefont{Guinea}},
  \bibinfo{author}{\bibfnamefont{A.}~\bibnamefont{Neto}}, \bibnamefont{and}
  \bibinfo{author}{\bibfnamefont{A.}~\bibnamefont{Lanzara}},
  \bibinfo{journal}{Nature Materials} \textbf{\bibinfo{volume}{6}},
  \bibinfo{pages}{770} (\bibinfo{year}{2007}).

\bibitem[{\citenamefont{Park et~al.}(2008{\natexlab{a}})\citenamefont{Park,
  Yang, Son, Cohen, and Louie}}]{PYSCL08}
\bibinfo{author}{\bibfnamefont{C.-H.} \bibnamefont{Park}},
  \bibinfo{author}{\bibfnamefont{L.}~\bibnamefont{Yang}},
  \bibinfo{author}{\bibfnamefont{Y.-W.} \bibnamefont{Son}},
  \bibinfo{author}{\bibfnamefont{M.~L.} \bibnamefont{Cohen}}, \bibnamefont{and}
  \bibinfo{author}{\bibfnamefont{S.~G.} \bibnamefont{Louie}},
  \bibinfo{journal}{Nature Physics} \textbf{\bibinfo{volume}{4}},
  \bibinfo{pages}{213} (\bibinfo{year}{2008}{\natexlab{a}}).

\bibitem[{\citenamefont{Park et~al.}(2008{\natexlab{b}})\citenamefont{Park,
  Yang, Son, Cohen, and Louie}}]{PYSCL08b}
\bibinfo{author}{\bibfnamefont{C.-H.} \bibnamefont{Park}},
  \bibinfo{author}{\bibfnamefont{L.}~\bibnamefont{Yang}},
  \bibinfo{author}{\bibfnamefont{Y.-W.} \bibnamefont{Son}},
  \bibinfo{author}{\bibfnamefont{M.~L.} \bibnamefont{Cohen}}, \bibnamefont{and}
  \bibinfo{author}{\bibfnamefont{S.~G.} \bibnamefont{Louie}},
  \bibinfo{journal}{Phys. Rev. Lett.} \textbf{\bibinfo{volume}{101}},
  \bibinfo{pages}{126804} (\bibinfo{year}{2008}{\natexlab{b}}).

\bibitem[{\citenamefont{Tiwari and Stroud}(2009)}]{TS09}
\bibinfo{author}{\bibfnamefont{R.~P.} \bibnamefont{Tiwari}} \bibnamefont{and}
  \bibinfo{author}{\bibfnamefont{D.}~\bibnamefont{Stroud}},
  \bibinfo{journal}{Phys. Rev. B} \textbf{\bibinfo{volume}{79}},
  \bibinfo{pages}{205435} (\bibinfo{year}{2009}).

\bibitem[{\citenamefont{Brey and Fertig}(2009)}]{BF09}
\bibinfo{author}{\bibfnamefont{L.}~\bibnamefont{Brey}} \bibnamefont{and}
  \bibinfo{author}{\bibfnamefont{H.~A.} \bibnamefont{Fertig}},
  \bibinfo{journal}{Phys. Rev. Lett.} \textbf{\bibinfo{volume}{103}},
  \bibinfo{pages}{046809} (\bibinfo{year}{2009}).

\bibitem[{\citenamefont{Barbier et~al.}(2010)\citenamefont{Barbier,
  Vasilopoulos, and Peeters}}]{BVP10}
\bibinfo{author}{\bibfnamefont{M.}~\bibnamefont{Barbier}},
  \bibinfo{author}{\bibfnamefont{P.}~\bibnamefont{Vasilopoulos}},
  \bibnamefont{and} \bibinfo{author}{\bibfnamefont{F.~M.}
  \bibnamefont{Peeters}} (\bibinfo{year}{2010}), \eprint{arXiv:1002.1442v1}.

\bibitem[{\citenamefont{Arovas et~al.}(2010)\citenamefont{Arovas, Brey, Fertig,
  Kim, and Ziegler}}]{ABFKZ10}
\bibinfo{author}{\bibfnamefont{D.~P.} \bibnamefont{Arovas}},
  \bibinfo{author}{\bibfnamefont{L.}~\bibnamefont{Brey}},
  \bibinfo{author}{\bibfnamefont{H.~A.} \bibnamefont{Fertig}},
  \bibinfo{author}{\bibfnamefont{E.-A.} \bibnamefont{Kim}}, \bibnamefont{and}
  \bibinfo{author}{\bibfnamefont{K.}~\bibnamefont{Ziegler}}
  (\bibinfo{year}{2010}), \eprint{arXiv:1002.3655}.

\bibitem[{\citenamefont{Snyman}(2009)}]{S09}
\bibinfo{author}{\bibfnamefont{I.}~\bibnamefont{Snyman}},
  \bibinfo{journal}{Phys. Rev. B} \textbf{\bibinfo{volume}{80}},
  \bibinfo{pages}{054303} (\bibinfo{year}{2009}).

\bibitem[{\citenamefont{Bliokh et~al.}(2009)\citenamefont{Bliokh, Freilikher,
  Savel'ev, and Nori}}]{BFSN09}
\bibinfo{author}{\bibfnamefont{Y.~P.} \bibnamefont{Bliokh}},
  \bibinfo{author}{\bibfnamefont{V.}~\bibnamefont{Freilikher}},
  \bibinfo{author}{\bibfnamefont{S.}~\bibnamefont{Savel'ev}}, \bibnamefont{and}
  \bibinfo{author}{\bibfnamefont{F.}~\bibnamefont{Nori}},
  \bibinfo{journal}{Phys. Rev. B} \textbf{\bibinfo{volume}{79}},
  \bibinfo{pages}{075123} (\bibinfo{year}{2009}).

\bibitem[{\citenamefont{Abedpour et~al.}(2009)\citenamefont{Abedpour,
  Esmailpour, Asgari, and Tabar}}]{AEAT09}
\bibinfo{author}{\bibfnamefont{N.}~\bibnamefont{Abedpour}},
  \bibinfo{author}{\bibfnamefont{A.}~\bibnamefont{Esmailpour}},
  \bibinfo{author}{\bibfnamefont{R.}~\bibnamefont{Asgari}}, \bibnamefont{and}
  \bibinfo{author}{\bibfnamefont{M.~R.} \bibnamefont{Tabar}},
  \bibinfo{journal}{Phys. Rev. B} \textbf{\bibinfo{volume}{79}},
  \bibinfo{pages}{165412} (\bibinfo{year}{2009}).

\bibitem[{\citenamefont{Shytov et~al.}(2009)\citenamefont{Shytov, Abanin, and
  Levitov}}]{SAL09}
\bibinfo{author}{\bibfnamefont{A.~V.} \bibnamefont{Shytov}},
  \bibinfo{author}{\bibfnamefont{D.~A.} \bibnamefont{Abanin}},
  \bibnamefont{and} \bibinfo{author}{\bibfnamefont{L.~S.}
  \bibnamefont{Levitov}}, \bibinfo{journal}{Phys. Rev. Lett.}
  \textbf{\bibinfo{volume}{103}}, \bibinfo{pages}{016806}
  (\bibinfo{year}{2009}).

\bibitem[{\citenamefont{Gibertini et~al.}(2009)\citenamefont{Gibertini, Singha,
  Pellegrini, Polini, Vignale, Pinczuk, Pfeiffer, and West}}]{Getal09}
\bibinfo{author}{\bibfnamefont{M.}~\bibnamefont{Gibertini}},
  \bibinfo{author}{\bibfnamefont{A.}~\bibnamefont{Singha}},
  \bibinfo{author}{\bibfnamefont{V.}~\bibnamefont{Pellegrini}},
  \bibinfo{author}{\bibfnamefont{M.}~\bibnamefont{Polini}},
  \bibinfo{author}{\bibfnamefont{G.}~\bibnamefont{Vignale}},
  \bibinfo{author}{\bibfnamefont{A.}~\bibnamefont{Pinczuk}},
  \bibinfo{author}{\bibfnamefont{L.~N.} \bibnamefont{Pfeiffer}},
  \bibnamefont{and} \bibinfo{author}{\bibfnamefont{K.~W.} \bibnamefont{West}},
  \bibinfo{journal}{Phys. Rev. B} \textbf{\bibinfo{volume}{79}},
  \bibinfo{pages}{241406} (\bibinfo{year}{2009}).

\bibitem[{\citenamefont{Rosales et~al.}(2009)\citenamefont{Rosales, Pacheco,
  Barticevic, {Le\'on}, {Latg\'e}, and Orellana}}]{Retal09}
\bibinfo{author}{\bibfnamefont{L.}~\bibnamefont{Rosales}},
  \bibinfo{author}{\bibfnamefont{M.}~\bibnamefont{Pacheco}},
  \bibinfo{author}{\bibfnamefont{Z.}~\bibnamefont{Barticevic}},
  \bibinfo{author}{\bibfnamefont{A.}~\bibnamefont{{Le\'on}}},
  \bibinfo{author}{\bibfnamefont{A.}~\bibnamefont{{Latg\'e}}},
  \bibnamefont{and} \bibinfo{author}{\bibfnamefont{P.~A.}
  \bibnamefont{Orellana}}, \bibinfo{journal}{Phys. Rev. B}
  \textbf{\bibinfo{volume}{80}}, \bibinfo{pages}{073402}
  (\bibinfo{year}{2009}).

\bibitem[{\citenamefont{Guinea et~al.}(2010)\citenamefont{Guinea, Katsnelson,
  and Geim}}]{GKG10}
\bibinfo{author}{\bibfnamefont{F.}~\bibnamefont{Guinea}},
  \bibinfo{author}{\bibfnamefont{M.~I.} \bibnamefont{Katsnelson}},
  \bibnamefont{and} \bibinfo{author}{\bibfnamefont{A.~K.} \bibnamefont{Geim}},
  \bibinfo{journal}{Nature Phys.} \textbf{\bibinfo{volume}{6}},
  \bibinfo{pages}{30} (\bibinfo{year}{2010}).

\bibitem[{\citenamefont{Park et~al.}(2009)\citenamefont{Park, Son, Yang, Cohen,
  and Louie}}]{PSYCL10}
\bibinfo{author}{\bibfnamefont{C.-H.} \bibnamefont{Park}},
  \bibinfo{author}{\bibfnamefont{Y.-W.} \bibnamefont{Son}},
  \bibinfo{author}{\bibfnamefont{L.}~\bibnamefont{Yang}},
  \bibinfo{author}{\bibfnamefont{M.~L.} \bibnamefont{Cohen}}, \bibnamefont{and}
  \bibinfo{author}{\bibfnamefont{S.~G.} \bibnamefont{Louie}},
  \bibinfo{journal}{Phys. Rev. Lett.} \textbf{\bibinfo{volume}{103}},
  \bibinfo{pages}{046808} (\bibinfo{year}{2009}).

\bibitem[{\citenamefont{da~{Silva-Ara\'ujo}
  et~al.}(2010)\citenamefont{da~{Silva-Ara\'ujo}, Chacham, and Nunes}}]{SCN10}
\bibinfo{author}{\bibfnamefont{J.}~\bibnamefont{da~{Silva-Ara\'ujo}}},
  \bibinfo{author}{\bibfnamefont{H.}~\bibnamefont{Chacham}}, \bibnamefont{and}
  \bibinfo{author}{\bibfnamefont{R.~W.} \bibnamefont{Nunes}},
  \bibinfo{journal}{Phys. Rev. B} \textbf{\bibinfo{volume}{81}},
  \bibinfo{pages}{193405} (\bibinfo{year}{2010}).

\bibitem[{\citenamefont{{Ma\~nes} et~al.}(2007)\citenamefont{{Ma\~nes}, Guinea,
  and Vozmediano}}]{MGV07}
\bibinfo{author}{\bibfnamefont{J.~L.} \bibnamefont{{Ma\~nes}}},
  \bibinfo{author}{\bibfnamefont{F.}~\bibnamefont{Guinea}}, \bibnamefont{and}
  \bibinfo{author}{\bibfnamefont{M.~A.~H.} \bibnamefont{Vozmediano}},
  \bibinfo{journal}{Phys. Rev. B} \textbf{\bibinfo{volume}{75}},
  \bibinfo{pages}{155424} (\bibinfo{year}{2007}).

\bibitem[{\citenamefont{Vozmediano et~al.}(2010)\citenamefont{Vozmediano,
  Katsnelson, and Guinea}}]{VKG10}
\bibinfo{author}{\bibfnamefont{M.~A.~H.} \bibnamefont{Vozmediano}},
  \bibinfo{author}{\bibfnamefont{M.~I.} \bibnamefont{Katsnelson}},
  \bibnamefont{and} \bibinfo{author}{\bibfnamefont{F.}~\bibnamefont{Guinea}}
  (\bibinfo{year}{2010}), \eprint{arXiv:1003.5179}.

\bibitem[{\citenamefont{Guinea et~al.}(2008)\citenamefont{Guinea, Horovitz, and
  Doussal}}]{GHL08}
\bibinfo{author}{\bibfnamefont{F.}~\bibnamefont{Guinea}},
  \bibinfo{author}{\bibfnamefont{B.}~\bibnamefont{Horovitz}}, \bibnamefont{and}
  \bibinfo{author}{\bibfnamefont{P.~L.} \bibnamefont{Doussal}},
  \bibinfo{journal}{Phys. Rev. B} \textbf{\bibinfo{volume}{77}},
  \bibinfo{pages}{205421} (\bibinfo{year}{2008}).

\bibitem[{\citenamefont{Haldane}(1988)}]{H88}
\bibinfo{author}{\bibfnamefont{F.~D.~M.} \bibnamefont{Haldane}},
  \bibinfo{journal}{Phys. Rev. Lett.} \textbf{\bibinfo{volume}{61}},
  \bibinfo{pages}{2015} (\bibinfo{year}{1988}).

\end{thebibliography}
\end{document}